\documentclass[12pt,letter]{article}
\usepackage{graphics}
\usepackage{graphicx}
\usepackage[latin1]{inputenc}
\usepackage{amsmath}
\usepackage{amssymb}

\newcommand{\Lyx}{L\kern-.1667em\lower.25em\hbox{y}\kern-.125emX\spacefactor1000}

\newcommand{\erfc}{\mathop{\text{erfc}}}
\newcommand{\erf}{\mathop{\text{erf}}}

\begin{document}
\bibliographystyle{plain} 
\pagestyle{plain} 
\pagenumbering{arabic}

\title{Statistical Behavior Of Domain Systems}
\author {Diego Luis Gonz\'alez\footnote{die-gon1@uniandes.edu.co}\\
        Gabriel T\'ellez\footnote{gtellez@uniandes.edu.co}\\
        Departamento de F\'{\i}sica, Universidad de Los Andes\\
        A.~A.~4976 Bogot\'a, Colombia.}

\date{}

\maketitle


\begin{abstract}
We study the statistical behavior of two out of equilibrium
systems. The first one is a quasi one-dimensional gas with two species
of particles under the action of an external field which drives each
species in opposite directions. The second one is a one-dimensional
spin system with nearest neighbor interactions also under the
influence of an external driving force. Both systems show a dynamical
scaling with domain formation. The statistical behavior of these
domains is compared with models based on the coalescing random walk
and the interacting random walk. We find that the scaling domain size
distribution of the gas and the spin systems is well fitted by the
Wigner surmise, which lead us to explore a possible connection between
these systems and the circular orthogonal ensemble of random
matrices. However, the study of the correlation function of the domain
edges, show that the statistical behavior of the domains in both gas
and spin systems, is not completely well described by circular
orthogonal ensemble, nor it is by other models proposed such as the
coalescing random walk and the interacting random walk. Nevertheless,
we find that a simple model of independent intervals describe more
closely the statistical behavior of the domains formed in these
systems.
\end{abstract} 



\section{Introduction}
In this article we study of the statistical behavior of two
non-equilibrium systems. The first one is a quasi one-dimensional gas
introduced in Ref.~\cite{mettetal}. There, the authors studied the
biased diffusion of two species in a fully periodic $2\times L$
rectangular lattice half filled with two equal number of two types of
particles (labeled by their charge $+$ or $-$). An infinite external
field drives the two species in opposite directions along the $x$ axis
(long axis). The only interaction between particles is an excluded
volume constraint, i.e., each lattice site can be occupied at most by
only one particle.

The second system is a one-dimensional spin system introduced in
Ref.~\cite{cornell}, where the authors consider a chain of $L$ Ising
spins with nearest neighbor ferromagnetic interaction $J$. The chain
is subject to spin-exchange dynamics with a driving force $E$ that
favors motion of up spins to the right over motion to the left.

Both systems evolve with formation of domains which grow in time. For
intermediate times where the size of the domains is much smaller than
the total size $L$ of the system, the domain size distribution exhibit a
dynamic scaling. In this work we will be interested in the statistical
properties of these domains in the scaling regime. 

For each system there are two kind of domains. For the gas system,
there are domains filled with particles and empty domains. For the spin
system there are domains of up spins and domains of down spins. We are
interested in the statistical properties of these domains regardless
of their type (filled or empty, up or down). Let us define
$P^{\left(n\right)}(S,t)$ as the probability density function that the
distance between the external borders of $n+1$ consecutive domains is
$S$ at time $t$. Let $\left\langle S\right\rangle$ be the average of
$S$ and the relative spacing between borders is $s=S/\left\langle
S\right\rangle$. The scaling probability density function is defined
as
\begin{equation}
p^{\left(n\right)}(s)=\left\langle S\right\rangle
P^{\left(n\right)}(s\left\langle S\right\rangle,t)
\,,
\end{equation} 
in particular $p^{\left(0\right)}(s)$ is the scaling nearest neighbor
edges distribution, i.e.~the domain size distribution. In the scaling
regime, the scaling probability density functions $p^{(n)}(s)$ do not
depend on the time $t$.

Numerical simulation of both systems performed in~\cite{mettetal,
cornell,spirin} show that their scaling domain size distribution function
is well fitted by
\begin{equation}\label{dgoe}
p^{\left(0\right)}(s)=\frac{\pi}{2}s\,e^{-\pi s^2/4}\,.
\end{equation} 

This distribution is known in the random matrices theory as the Wigner
surmise and it describes the spacing distribution between eigenvalues
in the Gaussian and circular orthogonal ensembles (GOE and COE
respectively)~\cite{mehta}. This fact suggest a possible connection
between the statistical behavior of the borders of domains in
non-equilibrium systems and the eigenvalues in random matrices in a
similar way as it occurs with the eigenvalues of the Gaussian
orthogonal/unitary ensemble and the vicious random walk
\cite{fisher,forrester,forrester1}.

The main objective of this paper is to explore this possible
connection between non-equilibrium domain systems and random matrices,
in particular the circular orthogonal ensemble. In
sections~\ref{sec:gas} and~\ref{sec:spin}, we summarize the main
results found in Ref.~\cite{mettetal} on the quasi one-dimensional
gas, and in Ref.~\cite{cornell,spirin} on the spin system. In
section~\ref{sec:rand-mat}, we recall some facts about the random
matrix theory, relevant for our purposes. In section~\ref{sec:res}, we
compare the statistical behavior of the domains for each system with
the statistical behavior of the eigenvalues of random matrices from
the circular orthogonal ensemble. Also, we will consider a simple
model of independent intervals, which turns out to better describe the
statistics of the domains in both the gas and the spin systems.

\section{Quasi one-dimensional gas}
\label{sec:gas}

This system was described briefly in the introduction, more details
are found in Ref.~\cite{mettetal}. The system evolves in time
according to the following dynamical rules:

\begin{enumerate}
\item $L$ particles are randomly inserted in a $2\times L$ rectangular
  lattice, $\frac{L}{2}$ particles $(+)$ and $\frac{L}{2}$ particles
  $(-)$, the remaining sites are empty. Periodic boundary conditions
  are imposed in both directions of the lattice. Let the $x$ axis be
  the long axis of length $L$.
\item Two neighbor sites are chosen at random. The contents of the
  sites are exchanged with probability $1$ if the neighbors sites are
  particle-hole, but if they are particle-particle the content are
  exchanged with probability $\gamma$. The exchanges which result in
  $+/-$ particles moving in the positive/negative $x$ direction are
  forbidden due to the action of the external field.
\item A time unit correspond to $2L$ attempts of exchange.
\end{enumerate}

\subsection{Qualitative results}

The system was studied by computer simulation in
Refs.~\cite{mettetal,Georgiev}. We also performed several simulations
of this system with a parameter $\gamma=0.1$. The following
qualitative results are found from the simulations:

\begin{itemize}
	
\item For low $\gamma$ values the system remains homogeneous,
  i.e., the system evolves without domain formation.
\item For high $\gamma$ values the system evolves as follows. For
  early times small domains form everywhere due to mutual obstruction
  of the opposite species. After a long time, the system settles into
  a non-equilibrium steady state (NESS) in which only one macroscopic
  domain survives, so that, its length is about $\frac{L}{2}$ and it
  contains almost no holes. In the low density region the traveler
  particles leak out of the domain at one end and later rejoin it at
  the other end. The charge distribution of this macroscopic domain is
  not trivial.
\item For intermediate times, when the average size of the clusters
  is much smaller than the size $L$ of the system, the system shows
  dynamical scaling.
\end{itemize}

\subsection{Quantitative results}

Following Ref.~\cite{mettetal}, we can map the quasi one-dimensional
lattice into a one-dimensional lattice. This approximation described
in Ref.~\cite{mettetal} is a coarse grained description. For any
configuration on the $2\times L$ lattice we construct an effective
one-dimensional one, with occupation numbers zero or one on a $L$
sites line, as follows. At each site $i$, we assign $0$ if there are
$5$ or less particles in the $10$ sites around it, including the
$i$-th column of the original lattice. We assign 1 otherwise. In the
coarse grained configuration a domain is a simple consecutive sequence
of ones and its size is just the length of this string.

Under this coarse grained description, the authors of
Ref.~\cite{mettetal} considered the ``residence distribution'' of the
clusters $\tilde P(S,t)$, defined as $\tilde P(S,t)= S P^{(0)} (S,t)$
in terms of the domain size distribution $P^{(0)}(S,t)$. Using the
residence distribution, they found~\cite{mettetal,Georgiev} that the
average length of domains $\bar{S}=\sum_{S} S \tilde P(S,t)/\sum_{S}
\tilde P(S,t)$ grows in time with an exponent of at least $0.6$. The
numerical results from the simulations of Ref.~\cite{mettetal}, show
that the scaling residence distribution $\tilde p(s)= \bar{S} \tilde
P( s \bar{S},t)$ is well fitted by
\begin{equation}
\tilde p(s)=\frac{32}{\pi^2} x^2 e^{-4x^2/\pi}
\,.
\end{equation}
Remembering that $\tilde p(s)$ is, by definition, proportional to $s
p^{(0)}(s)$, this implies that the scaling domain size distribution
$p^{(0)}(s)$ is properly fitted by the Wigner surmise for the
orthogonal ensembles of random matrices, Eq.~(\ref{dgoe}). We
performed our own simulations to confirm this, the results will be
shown in section~\ref{sec:res}.

In Ref.~\cite{mettetal}, the authors proposed an approximate model to
describe this system, in the coarse grained description, using the
coalescing random walk (CRW) where the particles execute independent
random walks, suffering a fusion reaction $\left(A+A\rightarrow
A\right)$ when two particles meet.  This approximation is useful
because the full hierarchy of correlations for this diffusion-limited
reaction has been solved, see
Ref.~\cite{ben-bur,Derrida-Godre,ben2,derrida2}. This approximation is
done under the following assumptions.
\begin{itemize}
\item The numbers of domains is equal to the number of
  walkers in the CRW.
\item The walkers describe random walks over a one dimensional ring
  with $N$ sites, in that way, $N$ its equal to the sum of the
  length of the domains.
\item The distances between walkers are equal to the length
  of the domains.
\item When two walkers meet they coalesce, representing the
  disappearance of a domain.
\item The traveling particles of the original model are neglected in
  the CRW description.
\end{itemize}
Under this approximation the spacing distribution is the same found as
in the coarse grained description, but the average length of domains
grow in time with an erroneous exponent of $0.5$.

In order to solve this problem, in Ref.~\cite{mettetal}, the authors
introduced the model of the interacting random walkers (IRW), where
neighboring walkers experience stronger attraction with decreasing
separation. The probability of a walker moving to the right is
\begin{equation}\label{eq:qr}
q_{r}=q_{s}\left(1+\left(\frac{0.06 \left\langle S\right\rangle}{l_{r}}\right)^{2}\right),
\end{equation}
where $l_{r}$ is the distance to its right nearest neighbor and
$q_{s}$ is the probability of staying. In the same way, the
probability of moving to the left is
\begin{equation}\label{eq:ql}
q_{l}=q_{s}\left(1+\left(\frac{0.06 \left\langle
S\right\rangle}{l_{l}}\right)^{2}\right)\,,
\end{equation}
where $l_{l}$ is the distance to its left nearest neighbor and
$q_{s}=1-q_{r}-q_{l}$. The factor $0.06 \left\langle S\right\rangle$
is chosen in Ref.~\cite{mettetal} in order to give the best fit to the
simulation data and the dependence with $l^{-2}$ is justified in
Ref.~\cite{mettetal}. In this model, $\bar{S}$ grows with an exponent
of $0.6$, see Ref.~\cite{mettetal}. The spacing distribution is still
given by~(\ref{dgoe}).

We performed a few simulations of this interacting random walk
changing the constant~0.06 in Eqs.~(\ref{eq:qr})
and~(\ref{eq:ql}). This has the effect of changing the growth exponent
of the domains, however the scaling domain size distribution
$p^{(0)}(s)$ remains unchanged, it is still given by the Wigner
surmise~(\ref{dgoe}).

\section{One-dimensional spin system}
\label{sec:spin}

In the introduction we briefly presented this system, for more details
see Ref.~\cite{cornell,spirin}. The lattice has a length $N$ with
$N\mu$ spins up (``$+$'') and $N(1-\mu)$ spins down (``$-$'') with
$0<\mu<1$. Periodic boundary conditions are imposed. The spin-flip
events are:
\begin{enumerate}
	\item $++--$ $\leftrightarrow$ $+-+-$ $\,\,\Delta=4J-E$.
	\item $--++$ $\leftrightarrow$ $-+-+$ $\,\,\Delta=4J+E$.
	\item $++-+$ $\leftrightarrow$ $+-++$ $\,\,\Delta=-E$.
	\item $-+--$ $\leftrightarrow$ $--+-$ $\,\,\Delta=-E$.
\end{enumerate}
where the transition probability rate for a process from left to right
is proportional to
$\frac{1}{2}\left(1-\tanh\left(\frac{\Delta}{2T}\right)\right)$.  The
constant $J$ is the nearest neighbor coupling between spins, $E$ is
the energy associated to an external field which drives the up
(``$+$'') spins to the right and the down (``$-$'') spins to the left,
and $T$ the thermal energy (temperature times Boltzmann constant).
 
As explained in~\cite{cornell}, the microscopic dynamics of the
lattice of spins may be mapped onto one for an array domain dynamics,
which provides a good approximation in the regime $T\ll E\ll J$. In
this approximation, domains of up spins move spontaneously to the
right, and domains of down spins move to the left. The rates for such
processes are independent of the domain size. The algorithm used for
the numerical simulation is the following:
\begin{enumerate}
	\item Set up a random array of alternating down and up spins, with $\mu N$ spins up and $\left(1-\mu\right)N$ spins down.
	\item Choose a domain at random.
	\item If the domain is down, move it to the left (i.e., reduce the size of its left neighbor by one and increase its right neighbor by one), otherwise move it to the right.
	\item If one of the neighbor domains is zero size, then the domain is removed.
	\item Update the clock by 1/number of domains.
	\item Repeat steps 2--6.
\end{enumerate}

This simplified system allows much better statistics than it would be
possible with the true microscopic system \cite{cornell}, and its
simulation is more easy to implement and it reproduces the same asymptotic
behavior.

\subsection{Qualitative results}
From the simulation, the following qualitative results are observed:
\begin{itemize}
\item For early times, little domains form everywhere.
\item At intermediate times, some domains disappear while
  other domains grow. In this time regime, the system shows
  dynamical scaling.
\item For later times, only two macroscopic domains remain,
  which move in opposite directions. The system falls in a
  non-equilibrium steady state.
\end{itemize}

\subsection{Quantitative results}

In Ref.~\cite{cornell}, the authors derive an analytical solution in
the case $\mu\rightarrow 0$, nevertheless we can derive another exact
solution for the dynamical domain model with $\mu=0.5$. Let
$P^{\left(0\right)}(n,m,t)$ be the frequency of finding a domain of
spin up with length $n$ and another domain of spin down with length
$m$ at time $t$. The master equation for $P^{\left(0\right)}(n,m,t)$
is
\begin{eqnarray}
\frac{dP^{\left(0\right)}(n,m,t)}{dt} & = & P^{\left(0\right)}(n+1,m,t)+P^{\left(0\right)}(n-1,m,t)-2P^{\left(0\right)}(n,m,t) \nonumber \\ 
& + & P^{\left(0\right)}(n,m+1,t)+P^{\left(0\right)}(n,m-1,t)-2P^{\left(0\right)}(n,m,t).\nonumber \\
\end{eqnarray} 
Let $\Delta x$ be the distance between nearest neighbor sites of the
lattice. In the continuous limit, $\Delta x\to 0$, the above
probability is a function of $S=n \Delta x$, $R=m \Delta x$. It
satisfies the master equation
\begin{equation}
\frac{dP^{\left(0\right)}(S,R,t)}{dt}=\frac{\partial^{2}P^{\left(0\right)}(S,R,t)}{\partial
S^{2}}+\frac{\partial^{2}P^{\left(0\right)}(S,R,t)}{\partial R^{2}}.
\end{equation}
If at $t=0$ all domains have length $S_0$, the above equation is subject to the initial condition $P^{\left(0\right)}(S,R,0)=\delta(S-S_{0})\delta(R-S_{0})$ and the boundary conditions $P^{\left(0\right)}(0,R,t)=0$ and $P^{\left(0\right)}(S,0,t)=0$. The appropriate scaling solution for this differential equation is
\begin{equation}
p^{\left(0\right)}(s,r)=\frac{\pi^2}{4}s\,r\,e^{-\frac{\pi s^2}{4}}e^{-\frac{\pi r^2}{4}},
\end{equation}
as a consequence, the length of domains of spin up and of domains of
spin down have the same probability distribution function found in the
quasi one-dimensional gas. The above equation can be written as the
product of two distributions with the form of the Wigner
surmise~(\ref{dgoe}).

The fact that the domain size distribution $p^{(0)}(s)$ for a domain of
up or down spins is well described by the Wigner surmise~(\ref{dgoe})
was also confirmed in Ref.~\cite{cornell,spirin} from the numerical simulation
results.
 
\section{Random matrices}
\label{sec:rand-mat}

Historically, the initial motivation to introduce random matrices was
to study the statistics of the energy levels of some quantum systems,
in nuclear physics initially, then for quantum systems when their
classical counterpart is chaotic, for a review see
Ref.~\cite{mehta}. Some local properties, such as the spacing between
consecutive high energy levels are well described by the random matrix
theory. For quantum chaotic systems with time-reversal invariance and
integral total angular momentum, the high energy levels, for a given
value of the spin and the parity, have spacing distributions and
correlations that are well described by the Gaussian orthogonal
ensemble (GOE) of random matrices, when the energy levels are rescaled
such that the average interspacing is unity. For large matrices, the
(properly rescaled) statistics of the eigenvalues of the GOE in the
bulk is the same as the one for the circular orthogonal ensemble
(COE). In this work we preferred to work with the circular orthogonal
ensemble rather than the Gaussian orthogonal ensemble since the former
one is periodic and that way we avoid boundary condition problems (in
the GOE the statistics of the eigenvalues near the edge are different
than in the bulk).

The COE is an statistical ensemble of $N\times N$ unitary symmetric
matrices $S$ invariant under the transformations $S\rightarrow W^{T} S
W$ where $W$ is any $N\times N$ unitary matrix. The eigenvalues of $S$
are of the form $e^{i\theta_{k}}$, with $k=1,\ldots,N$. The probability
density function for the eigenvalues is
\begin{equation}
  P_{N\beta}(\theta_1,\ldots,\theta_N)= 
  \frac{1}{Z_{N\beta}} \prod_{1\leq k < j\leq N} 
    \left|e^{i\theta_{k}}-e^{i\theta_{j}}\right|^{\beta}
\end{equation}
with $\beta=1$. The partition function is $Z_{N\beta}=(2\pi)^{N}
\Gamma(1+\beta N/2)/(\Gamma(1+\beta/2))^N$~\cite{mehta}.  It should be
noted that the eigenvalue probability density function of the COE is
equal to the Boltzmann factor of an equilibrium system of particles
moving on a circle and interacting through a log-potential, the
two-dimensional Coulomb potential: 
\begin{equation}
  P_{N\beta}(\theta_1,\ldots,\theta_N)=
  \frac{1}{Z_{N\beta}} e^{-\beta V_{N}(\theta_1,\ldots,\theta_N)}
\end{equation}
with 
\begin{equation}
  \label{eq:log-pot}
  V_{N}(\theta_1,\ldots,\theta_N)=
  \sum_{1\leq k < j\leq N} 
  -\ln\left|e^{i\theta_{k}}-e^{i\theta_{j}}\right| 
  \,.
\end{equation}

Up to an additive constant, the log-potential is scale invariant, thus
in these log-gases the density plays only the trivial role of fixing
the length scale of the problem. In the other systems we study here,
it is important to rescale the lengths in order to have the
interparticle average spacing equal to one. For the non-equilibrium
systems, this has been possible due to the fact that they exhibit a
scaling regime.

The COE is a homogeneous system, the density of eigenvalues is
constant on the circle, equal to $2\pi/N$. It is convenient to rescale
the lengths such that the average spacing between eigenvalues is
unity. Let $x_{k}=N\theta_{k}/(2\pi)$ be the rescaled eigenvalues
(note a small abuse of language here, since the eigenvalues are really
$e^{i\theta_k}$ not $\theta_k$). Analytic expressions for the
eigenvalue spacing distributions are known, however they are
complicated, see Ref.~\cite{mehta}. It turns out that a very accurate
approximate expression for the nearest neighbor spacing distribution
$p^{(0)}(s)$ is given by the Wigner surmise~(\ref{dgoe}). In the limit
of large matrices, $N\to\infty$, the correlation function between two
(rescaled) eigenvalues $x$ and $x'$ is given by
\begin{equation}
  g(r)=\left[\int_{r}^{\infty} s(r')\,dr'\right]
  \frac{ds}{dr}(r)+\left[s(r)\right]^2
\end{equation}
with $r=|x-x'|$ and $s(r)=\sin(\pi r)/(\pi r)$.

\section{Results and discussion}
\label{sec:res}

\subsection{The nearest neighbor spacing distribution and the Wigner surmise}

We performed simulations for the gas, the spin, the CRW, the IRW and
the circular orthogonal ensemble (COE) of random matrices with the
following parameters. For the gas, the spin and the CRW systems, we
used a lattice with 1000 sites, and for the IRW the lattice had 500
sites. The random matrix simulations were performed with
200$\times$200 matrices. For the non-equilibrium systems, the
simulations were carried out long enough to reach the scaling
regime. The initial density of particles in the CRW and the IRW was
$1/2$ and $2/3$ respectively. In the scaling regime, we build
histograms for the domain sizes (i.e.~the nearest neighbor spacings
for the CRW, IRW and COE). The data to build the histograms was taken
at three different times for each system in order to verify the
existence of a proper scaling regime. These times were $T=1000$,
$T=1500$ and $T=2000$ for the gas system; $T=10$, $T=18$ and $T=34$
for the spin system (the time unit of each system was defined in the
previous sections); $T=50$, $T=100$ and $T=200$ for the CRW, and
$T=100$, $T=150$ and $T=200$ for the IRW. For the IRW and the CRW
systems, the Monte Carlo time unit corresponds to a trial to move all
the particles of the system. To have appropriate statistics, we
performed 20000 simulations (realizations) of the gas, the spin and
the COE systems, and 50000 realizations of the CRW and the IRW
systems.

\begin{figure} [!h]
\begin{center}
\includegraphics[scale=0.8]{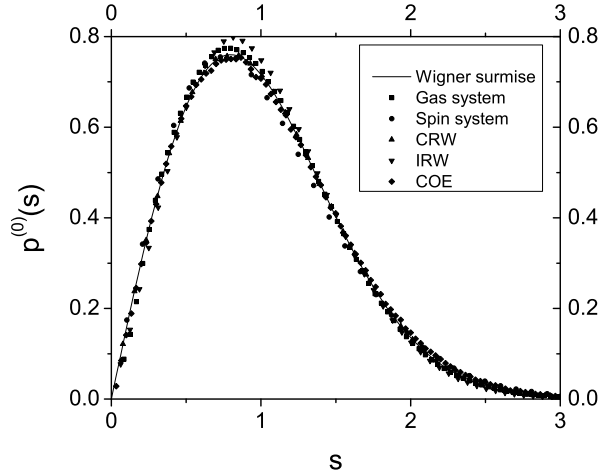}
\end{center}
\caption{Nearest neighbor spacing distributions for the gas, spin,
IRW, CRW and COE and Wigner surmise.}
\label{g0}
\end{figure}

Figure~\ref{g0} shows the scaling domain size distribution
$p^{(0)}(s)$ for the gas and the spin systems, obtained from the
simulations. It is compared with the scaling nearest neighbor
distribution of random walkers in the IRW and the CRW, and the scaled
nearest neighbor distribution of eigenvalues of the circular
orthogonal ensemble, which we also obtained from simulation of these
systems. We also present the Wigner surmise, Eq.~(\ref{dgoe}), which
is know to numerically reproduce accurately the nearest neighbor
distribution of eigenvalues of the circular orthogonal ensemble (it is
exact for 2$\times$2 matrices).

As we can see in figure~\ref{g0}, the Wigner surmise reproduces
correctly the nearest neighbor distribution
$p^{\left(0\right)}\left(s\right)$ for the gas, the spin, the IRW, the
CRW and the COE systems. This suggest a possible connection between the
random matrices and the out of equilibrium systems.

Motivated by this possible link between the non-equilibrium
statistical systems and the random matrix theory (i.e.~the equilibrium
log-gas system), we explore the possibility if not only the nearest
neighbor distribution is the same for all these systems, but also the
other spacing distributions $p^{\left(n\right)}(s)$ for $n>0$ and the
pair correlation function $g(r)$. The approach here is to consider the
edges of the domains in the gas and spin system as fundamental
entities, and to study their spacing distributions and correlation
function, in the scaling regime. With this study we wish to know if
the similitude between these systems goes beyond the nearest neighbor
distribution, or if it is simple a coincidence that the nearest
neighbor distribution is the same for all systems. This also can give
an indication about how much information on each system is contained
in the nearest neighbor distribution.

From our simulations, we computed the other spacing distributions
$p^{(n)}(s)$ and the correlation function $g(r)$ of all these systems
in order to compare them. The results are presented in the following
subsections.

\subsection{The quasi one-dimensional gas and its approximate models:
  the IRW and the CRW}

\begin{figure} [!h]
\begin{center}
\includegraphics[scale=0.8]{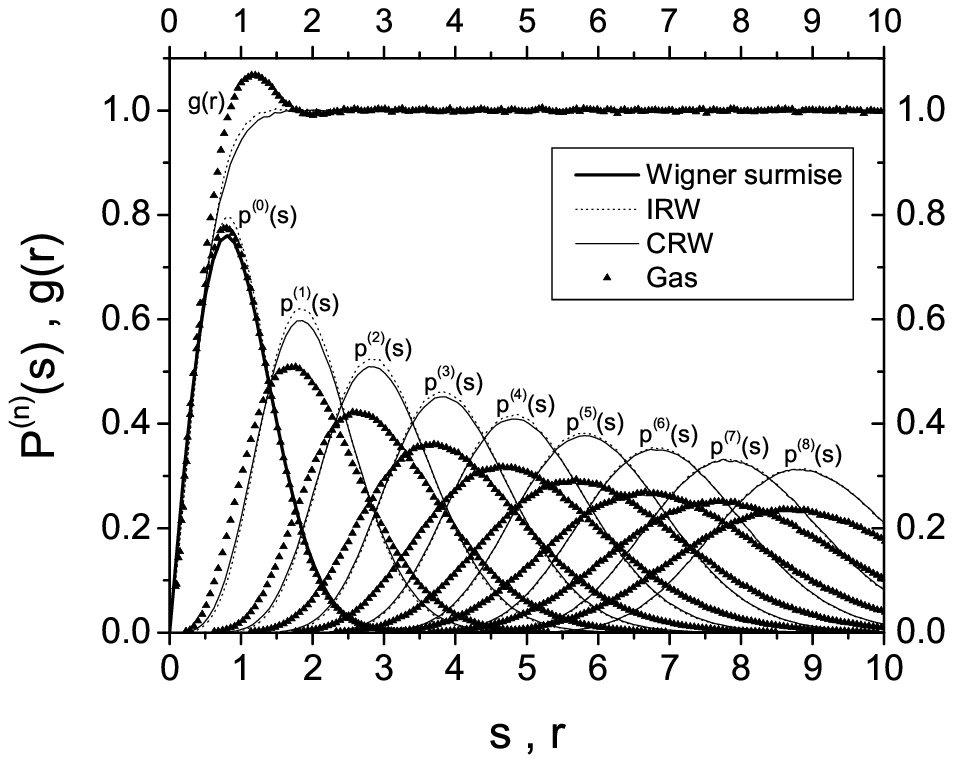}
\end{center}
\caption{Spacing distributions and the pair correlation function in
the gas system, IRW and CRW for a lattice with $N=500$ and $N=1000$
sites respectively.}
\label{g1}
\end{figure}

In Ref.~\cite{mettetal}, the domain size distribution of the quasi
one-dimensional gas was described by the approximate models of random
walkers, the IRW and CRW. We now compare the other spacing
distribution functions and the correlation function. The results are
shown in figure~\ref{g1}. The CRW and IRW reproduce correctly the
nearest neighbor distribution $p^{\left(0\right)}\left(s\right)$. But,
both the IRW and the CRW fail describing the other spacing
distribution functions $p^{\left(n\right)}\left(s\right)$ with $n>0$,
and, as consequence, they also fail describing the two-point
correlation function $g\left(r\right)$, since the correlation function
can be expressed as
 \begin{equation}\label{pkycdepares1}
g(r)=\sum^{\infty}_{n=0}p^{\left(n\right)}(r)
\,.
\end{equation}
We notice in particular the fact that the correlation function of the
gas system shows an oscillation around $r=1$ (and probably more for
higher values of $r$). On the contrary, the CRW and IRW models show no
oscillation in the correlation function $g(r)$. For these systems
$g(r)$ seems to be a monotonic function.

The spacing distributions $p^{(n)}(s)$ for $n\geq 1$ of the IRW and
CRW are somehow similar, but show higher maxima values than those of
the gas system. Furthermore, the maxima of the IRW and the CRW are
located at higher values of $s$ than the corresponding ones of the gas
system.

We conclude that the CRW and IRW systems are not statistically
equivalent to the quasi one-dimensional gas, although these models
were proposed in Ref.~\cite{mettetal} to describe (approximately) the
quasi one-dimensional gas.

\subsection{The gas and spin systems and the circular orthogonal
  random matrix ensemble (COE)}

\begin{figure} [!h]
\begin{center}
\includegraphics[scale=0.8]{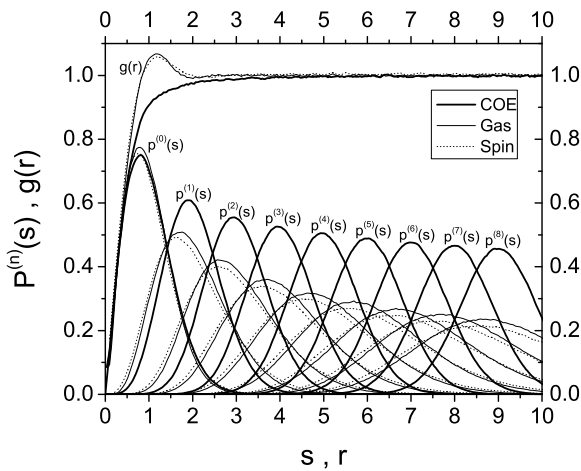}
\end{center}
\caption{Comparison between the gas, spin system ($\mu=0.5$) and COE for
  a lattice with $N=1000$ sites. In the COE we took $20000$ matrices
  of $200\times200$.}
\label{g3}
\end{figure}

Now, in figure~\ref{g3}, we compare the spacing distributions
$p^{(n)}(s)$ and the correlation function $g(r)$ of both the gas and
spin systems with the ones for the random matrix COE.

The correlation functions $g(r)$ of the gas and the spin system seem
to be very similar, almost identical, with only small differences of a
few percents in relative difference. They both exhibit the oscillation
near $r=1$ mentioned above. On the contrary, the correlation function
for the COE does not show any oscillation, and it is very different
from the correlation function of the gas and spin system.

The spacing distributions $p^{(n)}(s)$ are somehow similar for the gas
and the spin system, with some small differences between them. But, in
any case, they differ much from the ones of the COE.  The distributions
$p^{(n)}(s)$, for $n\geq 1$, are more ``disperse'' in the gas and spin
systems than the circular orthogonal ensemble, i.e.~they have a width
at half height larger than the ones for the COE. Also, the maxima for
each distribution of the gas and spin system are located at smaller
values of $s$ than for the COE. The maxima values are also smaller for
the gas and spin system than for the COE.

It is possible to conclude that the gas and spin systems have a
similar statistical behavior but this behavior is different from the
COE. They only coincide in the nearest neighbor distribution.

\begin{figure} [!h]
\begin{center}
\includegraphics[scale=0.8]{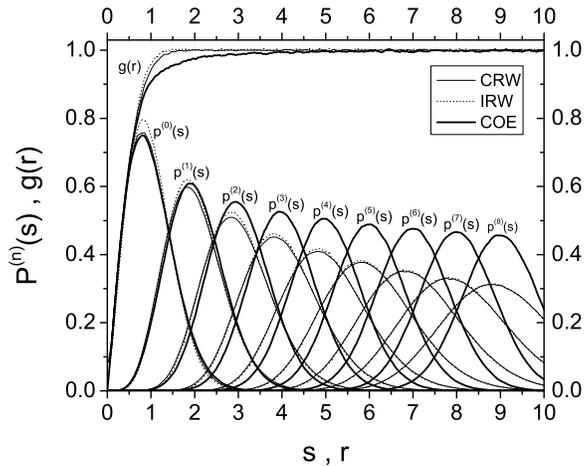}
\end{center}
\caption{Comparison between the CRW, IRW and COE for a lattice with
$N=1000$ sites and in COE we take $20000$ matrices of $200\times200$.}
\label{g4}
\end{figure}

In the previous subsection, we have shown that the CRW and the IRW
systems do not describe properly the statistics of the spacing
distributions of the gas system. But do they have a similar behavior
as the one of the COE? In figure~\ref{g4}, we compare the correlation
functions and spacing distribution functions of both the CRW and IRW
systems with COE.

The correlation function $g(r)$ of the CRW and IRW differs from the one
of the COE. Although the three correlation functions are monotonous
(they have no oscillations), around $1\lesssim r \lesssim 2$, the
correlation function of the COE is smaller than the one of the IRW and
the CRW. The first two distribution functions $p^{(0)}(s)$ and
$p^{(1)}(s)$ of the three systems, IRW, CRW and COE, are very
similar. However from $n\geq 2$, the COE spacing distributions
$p^{(n)}(s)$ start to differ from the corresponding ones of the CRW
and the IRW. The spacing distributions of the CRW and the IRW seem
more ``disperse'' (as defined above) than the ones of the COE. The
maxima values for the spacing distributions in the COE are larger than
for the IRW and CRW. Also, these maxima are located at smaller values
of $s$ for the IRW and CRW than for the COE.

We conclude that the CRW and the IRW systems are not statistically
equivalent to the COE. On the other hand the CRW and the IRW show very
similar spacing distribution functions and correlation functions. Thus,
another important conclusion is that the interaction proposed
in Ref.~\cite{mettetal} for the IRW do not change considerably the
statistical behavior of the domains of the system from the CRW in the
scaling regime. The interaction between neighbors, Eqs.~(\ref{eq:qr})
and~(\ref{eq:ql}), change the growth exponent of the domains. But,
once rescaled, the statistics of the walker spacing distributions and
correlations of the CRW and IRW are very similar.

\subsection{The independent interval approximation (IIA)}

The results presented above show that the statistics of the edges of
the domains of the out of equilibrium gas and spin systems, properly
rescaled in the scaling regime, are not described by the COE, thus are
not described by an equilibrium statistical system of particles with a
log-potential interaction. Furthermore, for the gas systems, the
approximate models of CRW and IRW do not reproduce correctly higher
order spacing distributions and the correlation function. In this
section, we propose a model that describe better the statistics of the
domains of the out of equilibrium systems in the scaling regime when
properly rescaled.

We have found numerical evidence that suggest that the statistical
behavior the domains of both the gas and spin systems can be fairly
well described by the independent interval approximation (IIA), where
the correlation between domains is neglected. With this
simplification, it is possible calculate the distributions
$p^{\left(n\right)}(s)$, for any $n\geq 1$, see
Refs.~\cite{alemany,derrida,majumdar, Krap-Naim-PRE,
Krap-Naim-JSP}. For example, for $n=1$, the probability to have two
domain edges separated by a distance $s$, knowing that there is one
edge in between at a distance $r$ from the first edge is
$p^{(0)}(r)p^{(0)}(s-r)$, if the domains are considered
independent. Therefore,
\begin{equation}
  \label{eq:p1-IIA}
  p^{\left(1\right)}(s)=\int^{s}_{0}dr\,
  p^{\left(0\right)}(r)p^{\left(0\right)}(s-r)
  \,.
\end{equation}  
It is useful to introduce here the Laplace transform of the
distribution functions
\begin{equation}
  \widetilde{p}^{(n)}(l)=\int_0^{\infty} p^{(n)}(s)e^{-ls}\,ds\,.
\end{equation}
Taking the Laplace transform of equation~(\ref{eq:p1-IIA}), we have
\begin{equation}
\label{eq:p1-lap}
\widetilde{p}^{\left(1\right)}(l)
=\left(\widetilde{p}^{\left(0\right)}(l)\right)^{2}
\,.
\end{equation}  
If the Wigner surmise~(\ref{dgoe}) is assumed for $p^{(0)}(s)$, its
Laplace transform is
\begin{equation}
  \label{eq:Laplace-Wigner}
  \widetilde{p}^{(0)}(l)=1-l
  e^{l^2/\pi}\erfc\left(\frac{l}{\pi}\right)
  \,,
\end{equation}
where $\erfc(z)=(2/\sqrt{\pi})\int_{z}^{\infty} e^{-t^2}$ is the
complementary Gaussian error function. Using the inverse Laplace
transform of expression~(\ref{eq:p1-lap}) or from a direct calculation
of Eq.~(\ref{eq:p1-IIA}), we find
\begin{equation}
p^{\left(1\right)}(s)=\frac{\pi}{16}e^{-\frac{\pi s^2}{4}}\left(4 s
+\sqrt{2}e^{\frac{\pi s^2}{8}}\left(-4+\pi
s^2\right)\mathrm{erf}\left(\frac{s}{2}\sqrt{\frac{\pi}{2}}\right)\right)
\,,
\end{equation}
with $\erf(z)=1-\erfc(z)$.

More generally, under this approximation, the Laplace transform of
$p^{(n)}(s)$ is simply
\begin{equation}
  \widetilde{p}^{(n)}(l)=\left(\widetilde{p}^{(0)}(l)\right)^n=
  \left(
  1-l e^{l^2/\pi}\erfc\left(\frac{l}{\pi}\right)
  \right)^n
  \,.
\end{equation}
Since the pair correlation function can be obtained from
Eq.~(\ref{pkycdepares1}), then $g(r)$ is given by the sum over
convolutions of $p^{\left(0\right)}(r)$, see Ref.~\cite{alemany}. In terms of the Laplace
transform $\widetilde{g}$ of the correlation function, we have
\begin{equation}\label{pkycdepares2}
\widetilde{g}(l)=
\sum^{\infty}_{k=1}\left(\widetilde{p}^{\left(0\right)}(l)\right)^k
=\frac{\widetilde{p}^{(0)}(l)}{1-\widetilde{p}^{(0)}(l)}
\,,
\end{equation}
with $\widetilde{p}^{(0)}(l)$ given by Eq.~(\ref{eq:Laplace-Wigner}).
We have
\begin{equation}
  \label{eq:g-r-IIA}
  \widetilde{g}(l)=\frac{1-le^{l^2/\pi}
    \erfc\left(\frac{l}{\sqrt{\pi}}\right)}{le^{l^2/\pi}
    \erfc\left(\frac{l}{\sqrt{\pi}}\right)}
  \,.
\end{equation}
The inverse Laplace transform of the above expression can be computed
numerically to obtain $g(r)$.

In this independent interval approximation, the joint probability
density function to find borders in positions $x_1,x_2,\cdots,x_N$ in
a line of length $L$ is given by
\begin{equation}
P_N(x_1,\cdots,x_N)=\frac{1}{Z_N}p^{(0)}(x_2-x_1)\cdots p^{(0)}(x_N-x_{N-1})p^{(0)}(x_{1}+L-x_{N}),
\end{equation}
in compact form
\begin{equation}\label{pjoin}
P_N(x_1,\cdots,x_N)=\frac{1}{Z_N}\prod^{N}_{i=1}p^{(0)}(x_{i+1}-x_i).
\end{equation}
where we have considered periodic boundary conditions and we defined
$x_{N+1}=x_1+L$. The partition function $Z_N$ is the normalization
constant
\begin{equation}
\label{eq:part-IIA}
Z_N=\int_{x_1<x_2<\ldots<x_N<x_1+L} dx_1\ldots dx_N\,
\prod^{N}_{i=1}p^{(0)}(x_{i+1}-x_i)\,.
\end{equation}
Comparing~(\ref{pjoin}) with a Boltzmann factor with an inverse
temperature $\beta=1$, we see that, under the independent interval approximation, the statistics of the domain edges is equivalent to a
system with $N$ particles which interact according to the potential
\begin{equation}\label{viia}
V_N(x_1,\cdots,x_N) =
\sum^{N}_{i=1}\left[
\frac{\pi}{4}\left(x_{i+1}-x_i\right)^2
-\ln\left(\frac{\pi}{2}\left(x_{i+1}-x_i\right)\right)
\right]\,.
\end{equation}
Thus, it is equivalent to a statistical equilibrium system of
particles on a circle interacting through a nearest neighbor pair
potential. The partition function and correlations can be computed by
means of the Laplace transform as explained in general for this kind
of systems in Ref.~\cite{salsburg}. We already computed the
correlation function $g(r)$ in Eqs.~(\ref{pkycdepares2})
and~(\ref{eq:g-r-IIA}). The partition function can be obtained as
follows. Doing a change of variable $x_{i}\to x_{i}-x_1$ in
Eq.~(\ref{eq:part-IIA}) we have
\begin{equation}
  Z_{N}=L\int_{0<x_2<\ldots<x_N<L}dx_2\cdots dx_N
  p^{(0)}(x_2)\left(\prod^{N}_{i=3}p^{(0)}(x_{i}-x_{i-1})\right)
  p^{(0)}(L-x_N)  \,.
\end{equation}
This is an $N$-fold convolution product of $p^{(0)}(x)$. Thus the
Laplace transform of the partition function is 
\begin{equation}
  \int_0^{\infty} e^{-lL}
  \frac{Z_N}{L}\,dL=(\widetilde{p}^{(0)}(l))^{N}
  =\left(
  1-l e^{l^2/\pi}\erfc\left(\frac{l}{\pi}\right)
  \right)^N
  \,.
\end{equation}

We recall that the circular orthogonal ensemble is equivalent to a
system of particles on a circle interacting through a log-potential,
see Eq.~(\ref{eq:log-pot}). From the equations~(\ref{viia})
and~(\ref{eq:log-pot}), it is clear the difference between the
independent domain approximation and the circular orthogonal ensemble,
in the first system there are only nearest neighbor interactions,
while in the second one every single particle interact with all the
other particles, the pair potential is not restricted to nearest
neighbors. Additionally, the functional form of both potentials is
different.

\begin{figure} [!h]
\begin{center}
\includegraphics[scale=0.8]{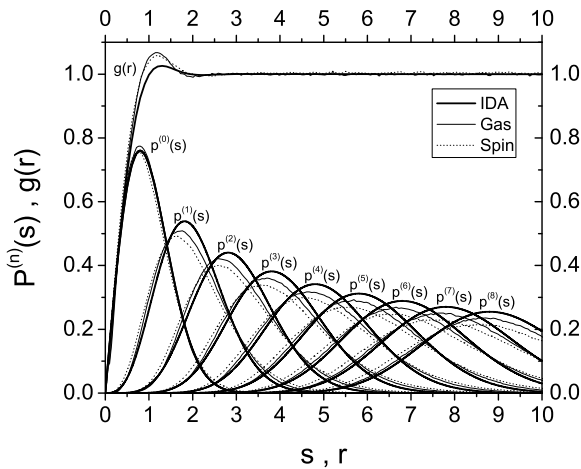}
\end{center}
\caption{Comparison between the gas, spin system and IIA for a lattice
with $N=1000$ sites.}
\label{g2}
\end{figure}

In figure~\ref{g2}, we compare the gas and spin system correlation
functions and spacing distributions with the theoretical predictions
from the independent interval approximation. We notice that this
approximation reproduce more closely the spacing distributions and the
correlation function. The IIA correlation function has an oscillatory
behavior near at $r=1$ as it occurs in the gas and spin
system. Although the predictions are not identical, the IIA gives a
good approximation for the correlation function and spacing
distributions with an error of a few percents (5\%) deviation from the
numerical results for the gas and the spin system.

The fact that the independent interval approximation reproduce much
better the correlation function and spacing distributions of the gas
and the spin system than the other approaches considered (CRW, IRW and
COE), suggest that in the gas and spin systems the domains are not
strongly correlated.

\section{Conclusion}
We studied two out of equilibrium systems, the gas and spin system,
which present a formation of growing domains. We studied the
statistical properties of the edges of the domains in the scaling
regime and compared them with several other models. Since in the
scaling regime the statistical properties of the edges of the domains
are time independent, when properly rescaled, we tried to find if
there exists an equilibrium statistical system which reproduce these
statistics. A first candidate was the circular orthogonal ensemble of
random matrices (COE), since the scaling size distribution of the
domains satisfies the Wigner surmise.

Studying the other spacing distributions $p^{(n)}(s)$ and the
correlation function, we found that only the nearest neighbor
distribution ($n=0$) coincides for the gas system, the spin system,
the COE and the models of coalescing and interacting random walkers
(CRW and IRW) which were introduced in Ref.~\cite{mettetal} to explain
the properties of the gas system. The distribution functions
$p^{(n)}(s)$ for $n\geq 1$ and the correlation function are different
for each of these systems. The nearest neighbor distribution for the
coalescing random walk is obtained by solving the diffusion equation
with an absorbing condition in $s=0$~\cite{mettetal, cornell}, and the
scaling solution turns out to be given by Eq.~(\ref{dgoe}). As a side
note, it is interesting to notice that, up to a normalization, this is
the wave function of the first excited state of the harmonic
oscillator, an analogy that has been used in the context of
persistence, see Ref.~\cite{majundar2,oerding}. Our work shows that it
is merely a coincidence that this nearest neighbor distribution turns
out to be the Wigner distribution from the random matrices theory.

We found a model which provides a better approximation for the gas and
the spin system domain statistics.  This model is the independent
interval approximation, even though in this approximation the
correlation between domains is not taken into account as it occurs in
interacting and coalescing random walk. The independent interval
approximation is statistically equivalent to an equilibrium system of
particles with only nearest neighbor interactions.

The fact that the Wigner surmise, Eq.~(\ref{dgoe}), reproduces
correctly the nearest neighbor spacing distribution of all the
different systems we considered here, but that the other spacing
distributions differ from system to system, seems to show that the
nearest neighbor distribution has some kind of universality while the
other distributions do not. However, it can also be interpreted as an
indication that the nearest neighbor distribution does not contain
enough information about the statistics of the system: several
different systems could share the same nearest neighbor distribution,
while the finer details and differences between them are contained in
the correlation function and the other spacing distributions. This is the 
most important result of this paper, because in many cases complex systems 
are mapped onto more simple systems with the only criteria of the nearest
neighbor distribution similitude. 

\section*{Acknowledgments}
The authors thank P.~J.~Forrester and F.~van Wijland for valuable
comments and observations. This work was partially supported by an
ECOS Nord/COLCIENCIAS action of French and Colombian cooperation.

\end{document}